\documentclass{article}
\usepackage{graphicx}
\usepackage{amsmath,amssymb}
\usepackage{graphics,color,array,calc,rotating,epsfig,psfrag}
\usepackage{hyperref}
\numberwithin{equation}{section}
\usepackage{cite}
\usepackage{bm}
\usepackage{dcolumn}
\usepackage{amsfonts}
\usepackage[mathscr]{eucal}
\def\be{\begin{equation}} \def\ee{\end{equation}}
\def\bea{\begin{eqnarray}} \def\eea{\end{eqnarray}}

\newcommand\prt{\partial}

\newcommand{\nn}{\nonumber}
\begin{document}
\baselineskip 18pt%
\begin{titlepage}
\vspace*{1mm}%
\hfill%
\vspace*{15mm}%
\hfill
\vbox{
    \halign{#\hfil         \cr
          } 
      }  
\vspace*{20mm}

\centerline{{\Large {\bf Central charges from Thermodynamics method in 3D gravity}}}
\vspace*{5mm}
\begin{center}
{Davood Mahdavian Yekta$^{a,}$\footnote{d.mahdavian@hsu.ac.ir}, Hanif Golchin$^{b,}$\footnote{h.golchin@uk.ac.ir}}\\
\vspace*{0.2cm}
{ $^{a}$Department of Physics, Hakim Sabzevari University, P.O. Box 397, Sabzevar, Iran\\
$^{b}$Faculty of Physics, Shahid Bahonar University of Kerman,
P.O. Box 76175, Kerman, Iran}\\
\vspace{1cm}
\end{center}

\begin{abstract}
In the context of guage/gravity duality, we investigate the central charges of a number of 2-dimensional conformal field theories (CFTs) that might live on the boundary of some 3-dimensional (3D) toy models of gravity, from the thermodynamics aspect of the black holes. For many black hole solutions, the entropy product of the inner Cauchy and outer event horizons is universal (mass independent). It is proposed that for these solutions, the central charges of the left- and right-moving sectors of the dual CFTs should be the same and one may read the central charges from the universal entropy product. This provides strong motivations for investigating this prescription for BTZ and Warped AdS$_{3}$ black holes in a number of 3D gravity theories and we will show that the proposal works truly. One striking result of our analysis is that if the entropy product is not universal in any theory of 3D gravity, then the left and right central charges are not equal.
\end{abstract}

\end{titlepage}

\section{Introduction}
One of the outstanding examples of simplified models that provides some insight into the full theory of (3+1)-dimensional (4D) general relativity is the (2+1)-dimensional (3D) toy model of gravity. Apart from being much simpler than the 4D case, it has been ‘solved’ in many different contexts and by different approaches \cite{Deser:1983tn,Witten:1988hc}.
According to the holographic picture \cite{Witten:1998qj} one may expect that for each sector of 3D gravity which is either asymptotically anti-de Sitter (AdS) or AdS-like, there exists a dual  (1+1)-dimensional conformal field theory (2D CFT). So, it will be interesting to investigate some properties of these 2D CFTs from different aspects such as black hole thermodynamics, asymptotic symmetry group, Lorentz Chern-Simons (CS) formalism, hidden conformal symmetry, counting near horizon microstate, and so on. Although the 3D Einstein-Hilbert (EH) gravity has no local degrees
of freedom, the higher order derivative deformations of EH gravity provide the theory with propagating degrees of freedom, i.e. 3D gravitons which could be also massive.

The first theory of this type was the topologically massive gravity (TMG) which was constructed by adding a cosmological constant and an odd-parity gravitational CS terms \cite{Deser:1981wh,Deser:1982vy}. The other parity preserving, ghosts-free, unitary and renormalizable model around the Minkowski ground state that includes certain curvature-squared terms, was the new massive gravity (NMG) \cite{Bergshoeff:2009hq,Bergshoeff:2009aq}. But the overall picture is changed when we go over to the AdS background, where various dynamical properties, such as unitarity and bulk to boundary behavior become more complex \cite{Liu:2009bk,Giribet:2009qz,Sinha:2010ai}. The general extension of NMG which contains infinite number of derivatives was also proposed in \cite{Gullu:2010pc} and is known as 3D Born-Infeld gravity. The 3D Einstein-Maxwell generalization of these theories were also proposed in refs. \cite{Moussa:2008sj,Ghodsi:2010gk,Ghodsi:2010ev}. There is another version of 3D gravity in AdS backgrounds which has one massive degrees of freedom as TMG and is unitary in both of the bulk and the CFT boundary in spite of the NMG known as minimal massive gravity (MMG) \cite{Bergshoeff:2014pca}. The quantization of these theories seems to give a richer structure than the EH theory and provides interesting toy models for higher-dimensional theories of quantum gravity.

On the other hand, the theory of GR includes  black hole solutions that obey the laws of thermodynamics and accompany by a macroscopic entropy \cite{Bardeen:1973gs,Bekenstein:1973ur}. However, we need a quantum theory of GR to produce this entropy from counting the micro-states \cite{Strominger:1996sh,Strominger:1997eq,Sen:1995in}, where the derivation
essentially relied on the presence of an AdS factor and on the application of the Cardy formula to its associated 2D conformal symmetry \cite{Brown:1986nw,Compere:2007in}. In this regard, 3D gravity is a candidate to study the quantum theory of GR that could also be appeared in the lower dimensional solutions of string theories \cite{Strominger:1997eq,Sfetsos:1997xs,Detournay:2010rh,Azeyanagi:2012zd} which embraces two practical features: the gauge/gravity duality or the AdS/CFT correspondence \cite{Maldacena:1997re,Gubser:1998bc}, and the black hole physics. Inspired by this duality, Strominger et al. proposed the Kerr/CFT correspondence \cite{Guica:2008mu} which states that the quantum fluctuations at the near-horizon of extremal Kerr black hole are identified with a 2D CFT which arises from the generators of asymptotic symmetry group of a certain boundary conditions imposed on the fluctuations. This duality is also generalized to many black holes in different gravity theories \cite{Lu:2008jk,Azeyanagi:2008kb,Chow:2008dp,Isono:2008kx,Ghodsi:2013soa,Ghodsi:2014fta,Sadeghian:2015hja}.

Even though most of the physical quantities of black holes are related to their outer event horizons, the inner Cauchy horizon mechanics are also important \cite{Ansorg:2009yi}. There are similar intensive thermodynamical quantities for the inner Cauchy horizon as outer event horizon, which can illuminate whether the first law of black hole thermodynamics works for them or not \cite{Cvetic:2010mn,Marolf:2011dj,Castro:2012av,Detournay:2012ug}.
It has been also long observed that the universality (mass-independence) of the entropy product of outer and inner horizons, i.e. $S^{+}S^{-}$, might play important roles in the holographic descriptions of black holes \cite{Cvetic:2010mn,Castro:2012av}.
For instance, in the case of general 4D and 5D multi-charged rotating black holes \cite{Larsen:1997ge,Cvetic:1997uw,Golchin:2019hlg}, the product $S_{+}S_{-}=4\pi^2 (N_{L}-N_{R})$, where $N_{L}$ and $N_{R}$ are interpreted as the levels of the left- and right-moving excitations of 2D CFT, is quantized and independent of the mass. In fact the right hand side is the level matching condition of the CFT, i.e. the requirement that the momentum along a compact spatial direction is quantized.

 A thermodynamics method of black hole/CFT correspondence has been proposed in \cite{Chen:2012mh,Chen:2012yd,Chen:2013rb} which shows that many universal quantities of the dual CFT could be constructed using the thermodynamics of the outer and inner horizons. Specifically, it has been observed in \cite{Chen:2012mh} that the entropy product is universal if the condition $T^+S^+=T^-S^-$ is satisfied and in this case, the central charges of left- and right-moving sectors of the dual CFT should be the same. Additionally, it is discussed \cite{Chen:2013rb} that one may read the central charges of the dual CFT from the universal entropy product.

The issue that we shall consider in this paper is the universality of entropy product for 3D BTZ \cite{Banados:1992wn} and spacelike Warped AdS$_3$ (WAdS$_3$) \cite{Moussa:2003fc,Anninos:2008fx} black holes and calculate the central charges of the dual CFT. This issue deserves further research along the lines that have already been proposed in \cite{Chen:2013rb}. This investigation has been done in the case of some higher dimensional charged rotating black holes in \cite{Chen:2013rb,Golchin:2020mkh}, and for 3D WAdS$_3$ black holes of TMG and NMG in \cite{Chen:2013aza,Zhang:2014wsa}. In particular, we investigate the entropy product and find the central charges of the dual CFT for these black holes in the case of NMG theory and two different extensions of it, that is, Maxwell-Born-Infeld (MBI) \cite{Ghodsi:2010ev} and generalized minimal massive gravity (GMMG) \cite{Setare:2014zea} theories. More specifically, it would be of interest to consider how this concept is applied for BTZ black holes in the exotic gravity theories \cite{Witten:1988hc,Townsend:2013ela,Ozkan:2018cxj}. One observation is the surprising fact that, in the case of solutions with universal entropy product, the central charges obtained from thermodynamics method are in agreement with the central charges that obtained from the other methods. In the solutions with mass dependent (non-universal) entropy product, we will show that the left and right central charges are not equal, which is consistent with the statements of thermodynamics method.

The structure of this paper is organized as follows: in section 2, we apply the  thermodynamics method for the BTZ and WAdS$_3$ black holes in NMG theory to find their dual CFT central charges. As a result, for these solutions the entropy product is universal and the central charges calculated from thermodynamics method are in agreement with those obtained from the other methods. In section 3, we consider the charged black holes in MBI gravity. The key feature being that, in spite of the presence of Maxwell CS term in the Lagrangian, this statement would be true for these solutions. In section 4, we investigate the GMMG and exotic gravity theories which embrace gravitational CS term and do not respect the universality.
Finally, the section 5 is dedicated to the concluding remarks.
\section{New Massive Gravity}
NMG is a unitary\footnote{In connection to ``bulk vs boundary'' clash, we should mention that the unitarity of the bulk gravity for $ m^2 \ell^2<\frac12$ leads to negative central charge which means the non-unitarity of the boundary theory. On the other hand, when the central charge is positive for unitary dual CFT, $m^2 \ell^2>\frac12$, the bulk theory is non-unitarity \cite{Li:2008dq,Liu:2009bk}.}, parity-preserving higher-derivative 3D theory of gravity \cite{Bergshoeff:2009hq} which is described by the following action
\be \label{LNMG} I_{NMG}=\frac{1}{2\kappa}\int d^3x \sqrt{-g} \left[R-2\Lambda_0-\frac{1}{m^2}\left(R_{\mu\nu}R^{\mu\nu}-\frac38 R^2\right)\right],\ee
where $m$ is a mass parameter, $\kappa\!=\!8\pi G$ and $\Lambda_0$ is the cosmological constant with the dimension of ({\it mass})$^2$. This theory admits BTZ as well as WAdS$_3$ black holes as solutions \cite{Clement:2009gq} and since we can write the Riemann tensor in terms of the Ricci tensors in 3D, it could be inherited to the Gauss-Bonnet gravity counterpart in higher dimensions \cite{Lovelock:1971yv}. In this section, we show that the entropy product of the outer event horizon and inner Cauchy horizon, is universal for these solutions and it is also possible to read the central charges of the dual 2D CFTs from this product according to the thermodynamics method in \cite{Chen:2013rb}.
\subsection{The BTZ black hole}
The black hole solution of Banados, Teitelbiom and Zanelli (BTZ) \cite{Banados:1992wn} has played a central role in studies of the physics of black holes, both classical and quantum. This is despite the fact that it is 3D, and that it has a negative cosmological constant. Its relative simplicity is part of the reason for its ubiquity, often providing simple (sometimes exact) results for a range of important phenomena from gravitational collapse to the scattering of quanta of the black hole. It even plays a direct role in studies of quantum aspects of classes of black holes in other dimensions, organizing the
microscopic accounting of the quantum degrees of freedom that underlie entropy \cite{Strominger:1997eq,Sfetsos:1997xs,Detournay:2010rh,Azeyanagi:2012zd}.

The BTZ black hole, which has the symmetries of AdS$_3$ geometry or $SL(2,R)_{L}\times SL(2,R)_{R}$ isometry, is mainly a solution of all 3D massive theories of gravity and is denoted by the line element in the form
\be\label{BTZ} ds^2=-N^2 dt^2+\frac{dr^2}{N^2}+r^2 \left(d\varphi+N^{\varphi} dt\right)^2,\ee
where
\be \label{func1} N^2=\frac{(r^2-r_{+}^2)(r^2-r_{-}^2)}{r^2 \ell^2}\,,\quad N^{\varphi}=\frac{r_{+} r_{-}}{r^2 \ell}\,,
\ee
with $\ell$ is the radius of AdS$_3$ space and $r_+ \,(r_-)$ refers to the outer (inner) horizon. The metric (\ref{BTZ}) is a solution for the equations of motion of the action (\ref{LNMG}) when
\be\label{p1} \Lambda_0=-\frac{4m^2 \ell^2+1}{4m^2 \ell^4}\,.\ee
The ADM mass and angular momentum of this solution can be calculated using the super angular momentum method proposed by Clement in \cite{Clement:1994sb} as
\be \label{mjbtz} M=\frac{r_{+}^2+r_{-}^2}{8G\ell^2}\left(1-\frac{1}{2m^2 \ell^2}\right),\quad J=\frac{r_{+}r_{-}}{4G\ell}\left(1-\frac{1}{2m^2 \ell^2}\right).\ee

It has been shown \cite{Saida:1999ec} that for a general 3D diffeomorphism invariant Lagrangian of the form $f(R_{\mu\nu},g_{\mu\nu})$, the entropy of BTZ black hole can be computed from the Wald formula \cite{Wald:1993nt} by using the expression $g^{\mu\nu} \frac{\prt f}{\prt R_{\mu\nu}}$. So the entropies are given by
\be\label{ebnmg}  S^{\pm}=\frac{\pi r_{\pm}}{2G}\left(1-\frac{1}{2m^2 \ell^2}\right),\ee
where $S^{\pm}$ are the entropy of outer ($r_{+}$) and inner ($r_{-}$) horizons, respectively. The Hawking temperatures and angular velocities of the solution are
\be\label{tbtz1} T^{\pm}=\frac{\kappa}{2\pi}\Big|_{r=r_{\pm}}=\frac{r_{+}^2-r_{-}^2}{2\pi\ell^2 r_{\pm}},\quad \Omega^{\pm}=N^{\varphi}\Big|_{r=r_{\pm}}=\frac{r_{\mp}}{\ell r_{\pm}}\,,\ee
where here, $\kappa$ is the surface gravity on the horizon.
It is straightforward to check that the above quantities satisfy the first law of black hole thermodynamics and Smarr-like formula as
\be\label{fsbtz} dM=\pm T^{\pm} dS^{\pm}+\Omega^{\pm} dJ\,, \qquad M=\pm \frac12 T^{\pm} S^{\pm}+\Omega^{\pm} J\,.\ee
Noting the relations (\ref{ebnmg}) and (\ref{tbtz1}), it is obvious that the equality $T^+S^+=T^-S^-$ is satisfied for the BTZ solution in NMG. According to \cite{Chen:2012mh}, this means that the entropy product of the outer and inner horizon is universal. In this case it is easy to check that
\be\label{prod10}
S^{+}S^{-}=\frac{\pi^2 \ell}{G}J\left(1-\frac{1}{2m^2 \ell^2}\right).
\ee

On the other hand, it is discussed \cite{Chen:2012mh, Chen:2013rb} that if the entropy product is universal, the central charges of the left- and right-moving sectors in the dual CFT must be the same and one may read the central charges from the entropy product as
\be \label{prod100}
c_{L}=c_{R}=6\, \frac{d}{dJ} \left(\frac{S^{+}S^{-}}{4\pi^2}\right).
\ee
Applying the above formula into the entropy product (\ref{prod10}) for the BTZ solution, one can find the central charge of the dual CFT
\be \label{CC1} c_{L}=c_{R}=\frac{3\ell}{2G}\left(1-\frac{1}{2m^2 \ell^2}\right),
\ee
which is in complete agreement with the results of the asymptotic symmetry group analysis done in \cite{Liu:2009kc,Ghodsi:2011ua,Yekta:2016fif}. Inserting (\ref{CC1}) in the Cardy formula, it is also possible to recover the entropies (\ref{ebnmg}) for the outer and inner horizons
\be
\label{cardy1} S^{\pm}=\frac{\pi^2}{3}(c_{L} T_{L}\pm c_{R}T_{R}),
\ee
where $T_{L,R}=\frac{1}{T_+}\pm\frac{1}{T_-}=\frac{r_{+}\pm r_{-}}{2\pi\ell}$\, are denoted in \cite{Maldacena:1998bw}.
It is also possible to find the energies of the left- and right-moving sectors of the dual CFT as
\be\label{en1} E_{L}=\frac{\pi^2}{3}c_{L} T_{L}^2=\frac{M\ell+J}{2}, \qquad E_{R}=\frac{\pi^2}{3}c_{R} T_{R}^2=\frac{M\ell-J}{2}\,.
\ee
\subsection{The WAdS$_3$ black hole}
Another solution for the NMG theory is the spacelike warped AdS$_3$ (WAdS$_3$) black hole that can be viewed as a fibration of the real line over AdS$_{2}$ with a constant warped factor. This breaks the $SL(2,R)_{L}\times SL(2,R)_{R}$ isometry group of AdS$_{3}$ down to $SL(2,R)\times U(1)$. The solution was found for the first time in the case of TMG theory in \cite{Moussa:2003fc,Anninos:2008fx} and thereafter by \cite{Clement:2009gq,Tonni:2010gb} for NMG. The metric describing this black hole in the ADM form is given by
\be\label{wads} \frac{ds^2}{\ell^2}=-N^2 dt^2+\frac{dr^2}{4N^2 K^2}+K^2 \left(d\varphi+N^{\varphi} dt\right)^2,\ee
where
\bea \label{func2} &4 K^2 N^2=(\nu^2+3)(r-r_{+})(r-r_{-})\,,\qquad 2K^2 N^{\varphi}=2\nu r-\sqrt{(\nu^2+3) r_{+} r_{-}},&\nn\\
&K^2=\frac{r}{4}\left[3(\nu^2-1)+(\nu^2+3)(r_{+}+ r_{-})-4\nu \sqrt{(\nu^2+3) r_{+} r_{-}}\right].&\eea
Here, the parameter $\nu$ is a dimensionless coupling that appears in the warped factor and the horizons are located at $r_{+}$ and $r_{-}$ where $g^{rr}$ as well as the determinant of the ($t,\varphi$) metric vanishes. The metric (\ref{wads}) is a solution for the equations of motion of the action (\ref{LNMG}) if we have the following relations between the constant parameters in NMG
\be\label{p2} \nu^2=\frac{2m^2\ell^2+3}{20}\,,\qquad \Lambda_0=\frac{4 m^4 \ell^4-468 m^2 \ell^2+189}{400 m^2 \ell^4}.\ee

The ADM mass and angular momentum of the WAdS$_3$ black hole calculated from a super angular momentum in Clement's approach are given in \cite{Tonni:2010gb,Ghodsi:2011ua} as
\bea \label{mjwads} M&\!\!\!=\!\!\!&\frac{\nu (\nu^2+3)}{G (20\nu^2-3)}\left(\nu(r_{+}+ r_{-})-\sqrt{(\nu^2+3) r_{+} r_{-}}\right),\nn\\
 J&\!\!\!=\!\!\!&\frac{\nu (\nu^2+3) \ell}{4G (20\nu^2-3)}\left[\left(\nu(r_{+}+ r_{-})-\sqrt{(\nu^2+3) r_{+} r_{-}}\right)^2-\nu^2 (r_{+}- r_{-})^2\right].\eea
 Also the Hawking temperatures and angular velocities related to the outer and inner horizons are given by
\be\label{twads} T^{\pm}=\frac{(\nu^2+3)}{4\pi\ell} \frac{r_{+}- r_{-}}{\left(2\nu r_{\pm}-\sqrt{(\nu^2+3) r_{+} r_{-}}\right)}\,,\qquad \Omega^{\pm}=\frac{2}{\ell \left(2\nu r_{\pm}-\sqrt{(\nu^2+3) r_{+} r_{-}}\right)}\,,\ee
 while the other thermodynamic quantities, i.e. the black hole entropies are
\be \label{ewads}
S^{\pm}=\frac{4\pi\nu^2 \ell}{G(20\nu^2-3)} \Big(2\nu r_{\pm}-\sqrt{(\nu^2+3) r_{+} r_{-}}\Big),
\ee
that could be computed from the Wald formula as in \cite{Clement:2009gq}. In the case of WAdS$_3$ black holes these parameters also satisfy the first law of black hole thermodynamics and Smarr-like formula as
\be\label{fswads} dM=\pm T^{\pm} dS^{\pm}+\Omega^{\pm} dJ\,, \qquad M=\pm T^{\pm} S^{\pm}+2\Omega^{\pm} J\,.\ee

According to the entropies and temperatures given in (\ref{ewads}) and (\ref{twads}), it is easy to check that the relation $T^+S^+=T^-S^-$ is satisfied for the WAdS$_3$ black hole, which means that the entropy product is universal. In this case one finds that
\be\label{prod11}
S^{+}S^{-}=\frac{(8\pi)^2\nu^3\ell}{G(20\nu^2-3)(\nu^2+3)} J\,.
\ee
Now, using (\ref{prod100}), one can read the central charges of the dual CFT from the entropy product. The result is
\be \label{CC2} c_{L}=c_{R}=\frac{96\nu^3\ell}{G(\nu^2+3)(20\nu^2-3)}\,.
\ee
The same result is obtained from the analysis of asymptotic symmetry group, as it is done in \cite{Yekta:2016jhg,Yekta:2016fif}. Considering the left and right temperatures as in \cite{Anninos:2008fx}
\be T_{L}=\frac{(\nu^2+3)}{8\pi} (r_{+}- r_{-}),\quad T_{R}=\frac{(\nu^2+3)}{8\pi \nu} \left(\nu(r_{+}+ r_{-})-\sqrt{(\nu^2+3) r_{+} r_{-}}\right),
 \ee
it is possible to find the energies of the left- and right-moving sectors of the dual CFT as
\be\label{en2} E_{L}=\frac{\pi^2}{3}c_{L} T_{L}^2=\frac{(20\nu^2-3)\ell G }{4\nu(\nu^2+3)}M^2, \quad E_{R}=\frac{\pi^2}{3}c_{R} T_{R}^2=\frac{(20\nu^2-3)\ell G }{4\nu(\nu^2+3)}M^2-J\,.
\ee
\section{Maxwell Born-Infeld gravity}
As in the NMG case, that a proper combination of quadratic curvature terms yields ghost-free action, one can define a remarkably simple gravitational Born-Infeld action that
extends NMG. The proposal of extending NMG to all orders was made in \cite{Gullu:2010pc} which is based on a 3D Born-Infeld type action. The other extension of NMG that includes the Maxwell theory also given in \cite{Ghodsi:2011ua} which is constructed by adding a Maxwell CS term to 3D Dirac-Born-Infeld action,  namely MBI, as
\be\label{DBINMG}
I_{MBI}=\frac{2m^2}{\kappa}\int d^3 x\Big[\sqrt{-|g_{\mu\nu}+\frac{\sigma}{m^2}\,G_{\mu\nu}+ a F_{\mu\nu}|}-(1+\frac{\Lambda_0}{2m^2})\sqrt{-| g_{\mu\nu}|}\,\Big]
+\frac{\mu}{2}\int d^3 x\,\epsilon^{\mu\nu\rho} A_{\mu}\partial_{\nu}A_{\rho}\,,
\ee
where $\sigma,a$, and $\mu$ are constant parameters, $G_{\mu\nu}$ and $F_{\mu\nu}=\partial_{\mu}A_{\nu}-\partial_{\nu}A_{\mu}$ are respectively the Einstein tensor and Maxwell field strength, and $\epsilon^{\mu\nu\rho}$ is the Levi-Civita tensor. For later convenience we choose $a^2\!\!=\!\!-\frac{\kappa}{2m^2}$ and $\sigma=-1$. Similar to NMG theory, $\kappa=8\pi G$ and $\Lambda_0$ are gravitational and cosmological constants. In this section we investigate universality for the  charged black holes in MBI theory which are asymptotically WAdS$_3$. Since the calculation of central charges of this theory from the asymptotic symmetry group is a challenging work, the thermodynamics method can be a proper tool to find them.

\subsection{Non-extremal charged black hole}
The physical charged black hole solutions for the equations of motion of (\ref{DBINMG}) which are free of closed time-like curves are described by the following ansatz \cite{Clement:1994sb,Ghodsi:2011ua}
\bea ds^2&\!\!\!=\!\!\!&-\beta^2\,\frac{\rho^2-\rho_0^2}{r^2}\,dt^2+r^2\big(d\varphi-\frac{\rho+(1-\beta^2)\,\omega}{r^2}\,dt\big)^2+\frac{1}{\beta^2 \zeta^2}\,\frac{d\rho^2}{\rho^2-\rho_0^2}\,,\nn\\
A&\!\!\!=\!\!\!&Q\left(2z\,dt-(\rho+2\omega z) d\varphi\right),
\eea
where $r^2=\rho^2+2\omega \rho+\omega^2 (1-\beta^2)+\frac{\beta^2 \rho_0^2}{1-\beta^2}$ and $\beta^2=1-2z$. To have a physical solution free of  closed time-like curves we must restrict ourselves to $0<\beta^2<1$ \cite{Moussa:2008sj}. $\zeta(\rho)$ is a scale factor allows for arbitrary reparametrizations of the radial coordinate $\rho$ and without loss of generality we choose $\zeta(\rho)=1$. The values of electric charge $Q$ and constant parameter $z$ are determined from the equations of motion. The background solution (vacuum) is obtained by inserting $\rho_0=\omega=0$. In this coordinate of space-time the black hole has two horizons at $\rho=\rho_0$ and $\rho=-\rho_0$ which are the outer and inner horizons, respectively. Further details are given in \cite{Ghodsi:2010ev,Ghodsi:2011ua} and references therein.

Solving the equations of motion gives the following relations for parameters
\be \label{p3}
z=\frac{16 (\mu^2-1)m^4+4(2\mu^2-3)m^2-3\mu^2}{48m^2},\quad \Lambda_0=\frac{(4m^2-1)\mu^2-6m^2 \mu+2m^2}{3\mu}.
 \ee
The physical parameters of the black hole are derived again by using the Clement's approach. In this way the angular momentum, mass, and electric charge are given by
\be\label{jmnonex}
J=4\pi(-\Xi_1\rho_0^2+\Xi_2\omega^2)\,,\quad M=8\pi\Xi_2\omega\,,\quad Q=\left[\frac{9-4m^2}{\kappa\mu^2(4m^2+3)}\right]^{\frac12},
\ee
where
\be \label{coeff}  \Xi_1=\frac{(1-2z)}{8\kappa\,z\,\mu},\quad \Xi_2= \frac{z(1-2z)}{2\kappa\mu}\,.
\ee
The entropies of outer and inner horizons are written in terms of parameters in (\ref{coeff}) as
 \be \label{enon}
S^{\pm}=\frac{16\pi^2}{(1-2z) \sqrt{2z}}(\pm 2z\Xi_1\rho_0+\Xi_2\omega),
\ee
and the Hawking temperatures, angular velocities and the electric potentials of the horizons are
 \be \label{term3}
T^{\pm}=\pm\frac{(1-2z)\,\rho_0}{A^{\pm}}\,,\quad\Omega^{\pm}=\frac{2\pi\sqrt{2z}}{A^{\pm}}\,,\quad \Phi^{\pm}=-\left(A_{t}+\Omega^{\pm} A_{\varphi}\right)=0\,,
\ee
where the area of event and Cauchy horizons in this coordinate are defined by
\be
A^{+}\equiv 2\pi r^{+}=\frac{2\pi}{\sqrt{2z}}\,(\rho_0+2\omega\,z), \quad A^{-}\equiv 2\pi r^{-}=\frac{2\pi}{\sqrt{2z}}\,(-\rho_0+2\omega\,z).
\ee

In the case of charged black holes we should also consider the contribution of the electric potential energy in the first law \cite{Ghodsi:2011ua}. So in this case the physical parameters of the non-extremal black hole satisfy in the first law and Smarr relations as follows
\be\label{fsnon} dM=\pm T^{\pm} dS^{\pm}+\Omega^{\pm} dJ+\Phi^{\pm} dQ\,, \qquad M=\pm \frac12 T^{\pm} S^{\pm}+\Omega^{\pm} J+\Phi^{\pm} Q\,.
\ee
Similar to the black holes studied in the previous section, the entropy product is universal for this solution. In this case on can find the universality relation as
\be \label{ss1}
S^+S^-=\frac{2\pi^2}{G\mu(1-2z)}J\,.
\ee
It is also possible to read the central charges of the dual CFT from product (\ref{ss1}) by using (\ref{prod100}), i.e.,
\be \label{CC3} c_{L}=c_{R}=\frac{3}{G\mu (1-2z)}.\ee
This is an exclusive result that could also be checked from other methods such as asymptotic symmetry group, CS Lagrangian formalism, phase space formalism and so forth. Noting the Cardy formula (\ref{cardy1}) and the left and right temperatures \cite{Ghodsi:2010ev} as
\be T_{L}=\frac{(1-2z)}{2\pi\sqrt{2z}}\rho_0 ,\quad T_{R}=\frac{(1-2z)}{2\pi\sqrt{2z}} 2\omega z\,, \ee
one can recover the entropies of the outer and inner horizons given in (\ref{term3}). Finally, the energies of the left- and right-moving sectors in the dual CFT are
\be\label{en3} E_{L}=\frac{\pi^2}{3}c_{L} T_{L}^2=\frac12 \omega M-J, \quad E_{R}=\frac{\pi^2}{3}c_{R} T_{R}^2=\frac12 \omega M\,.\ee

It is worth mentioning that in any diffeomorphism invariant theory of 3D gravity, just like the ones considered in the previous sections, the entropy products are independent of any electric or magnetic charges and only depend on the angular momentum of the black holes. So, the universality (mass independent) is true and one can calculate the central charges from the formula (\ref{prod100}) in the context of thermodynamics method. However, in dimensions $D\geq 4$, there are solutions that this product is a multi-charge function as $S^{+}S^{-}=\mathcal{F}(N_i)$. According to the thermodynamics method in the generic case \cite{Chen:2013rb}, one can read the central charges from the formula
\be
c_i=6 \frac{d}{dN_i} \big(\frac{S^{+}S^{-}}{4\pi^2}\big)\,,
\ee
where $N_i$s are the conserved charges of the solution like angular momenta, electric and magnetic charges. For instance, the Kerr-Newman black holes have $S^+S^-=4\pi^2 \big(J^2+\frac{Q^4}{4}\big)$ and the central charges of the dual CFTs should be calculated in the corresponding pictures \cite{Chow:2008dp} as $c_J=12J$ and $c_Q=6Q^3$.

\section{Entropy product in 3D gravity black holes}
So far, we have checked that if the entropy product of the outer and inner horizons for the 3D black hole solutions is universal, it is possible to read the central charges of the dual CFT from the entropy product according to the analysis of \cite{Chen:2012mh, Chen:2013rb}. In other words, the universality plays the key role in this analysis. In this section, we consider the BTZ and WAdS$_3$ black holes in some other 3D gravity theories and find the entropy product law for them.
\subsection{Generalized Minimal Massive Gravity}
In the context of AdS/CFT, the bulk/boundary clash is the major confinement for the initial 3D models such as TMG or NMG to be regarded as a good example of quantum gravity. It is impossible to arrange for both central charges of the asymptotic conformal symmetry algebra to be positive while also arranging for the bulk mode to have positive energy. To remedy this issue, it has been suggested another 3D toy model of gravity, dubbed ``minimal massive gravity'' (MMG) \cite{Bergshoeff:2014pca}, that has a single massive degree of freedom (like TMG) that is
unitary in the bulk and gives rise to a unitary CFT on the boundary. MMG is inherently described by a set of equations of motion that do not come from the variation of an action with respect to the metric as a dynamical field, however working within the ``Chern-Simons-like'' formulation \cite{Hohm:2012vh} one can define a 3-form Lagrangian that produces exactly the proposed MMG equations of motion.  We can construct higher-derivative extensions of 3D gravity in this formalism \cite{Afshar:2014ffa}. One of the extensions, dubbed generalized minimal massive gravity (GMMG), was obtained by unifying NMG with MMG in \cite{Setare:2014zea}.

The 3-form Lagrangian of GMMG is given by
\bea\label{GMMG} L_{GMMG}&\!\!\!=\!\!\!&-\sigma e\cdot R+\frac{\Lambda_0}{6} \,e\cdot e\times e+h\cdot T(\omega)+\frac{1}{2\mu} \left(\omega\cdot d\omega+\frac13 \omega\cdot\omega\times \omega \right)\nn\\
&&+\frac{\alpha}{2} e\cdot h\times h-\frac{1}{2m^2}\left(f\cdot R+\frac12 e\cdot f\times f\right).
\eea
The Lagrangian is parameterized by two dimensionless parameters $\sigma$ and $\alpha$, and a cosmological term $\Lambda_0$. It also contains two mass dimension constant $\mu$ and $m$.
Varying the above Lagrangian with respect to 1-form fields $e,\,\omega, \,f$, and $h$, one finds the following  equations of motion
\bea \label{eom1}
&-\sigma R(\omega)+\frac{\Lambda_0}{2} e \times e+D(\omega)h-\frac{1}{2m^2} f\times f+\frac{\alpha}{2}h\times h=0,&\\
&R(\omega)+\mu \,e\times h-\sigma \mu T(\omega)-\frac{\mu}{m^2} (df+\omega\times f)=0,&\\
&R(\omega)+e\times f=0,&\\
&T(\omega)+\alpha e\times h=0,& \eea
where the covariant exterior derivative, locally Lorentz covariant torsion and curvature 2-forms are respectively given by
\be \label{dtr}
D(\omega) h = dh + \omega \times h,\quad T(\omega)=de+\omega\times e,\quad R(\omega)=d\omega+\frac12\omega\times\omega.\ee
The equations of motion (\ref{eom1}) can also be recast in the original metric proposition \cite{Bergshoeff:2014pca} as follows
\be\label{eom2} \bar{\sigma} G_{\mu\nu}+\bar{\Lambda}_0 g_{\mu\nu}+\frac{1}{\mu} C_{\mu\nu}+\frac{\gamma}{\mu^2} J_{\mu\nu}+\frac{s}{2m^2}K_{\mu\nu}=0,\ee
where we can find the constant parameters $\bar{\sigma}, \bar{\Lambda}_0, \gamma$, and $s$ in terms of parameters in (\ref{GMMG}). We refer the reader to \cite{Setare:2014zea} for details of conventions.
\subsubsection{The BTZ black hole}
It has been shown in \cite{Setare:2015vea} that the BTZ black hole given in (\ref{BTZ}) is a solution for the GMMG field equations. The mass and angular momentum of the black hole are given by
\bea\label{jmbtzmmg} M&\!\!\!=\!\!\!& \left(\sigma+\frac{\gamma}{2\mu^2 \ell^2}+\frac{\eta}{2m^2\ell^2}\right)\frac{r_{+}^2+r_{-}^2}{8G\ell^2}-\frac{r_{+}r_{-}}{4G\mu\ell^2}\,\nn\\
J&\!\!\!=\!\!\!& \left(\sigma+\frac{\gamma}{2\mu^2 \ell^2}+\frac{\eta}{2m^2\ell^2}\right)\frac{r_{+}r_{-}}{4G\mu\ell^2}-\frac{r_{+}^2+r_{-}^2}{8G\ell^2}\,,
\eea
and the entropies of horizons are
\be\label{ebtzmmg}
S^{\pm}=\frac{\pi r_{\pm}}{2G}\left[\left(\sigma+\frac{\gamma}{2\mu^2 \ell^2}+\frac{\eta}{2m^2\ell^2}\right)-\frac{r_{\mp}}{\mu \ell r_{\pm}}\right].
\ee

One can check that the entropy product is not universal for this solution and it depends to the mass as
\be\label{prod40}
S^{+}S^{-}=\frac{\pi^2 \ell}{G}\left[\left(\sigma+\frac{\gamma}{2\mu^2 \ell^2}+\frac{\eta}{2m^2\ell^2}\right) J-\frac{1}{\mu } M\right].
\ee
In this case, according to the analysis of \cite{Chen:2012mh, Chen:2013rb}, we expect that the central charges of the left- and right-moving sectors in the dual CFT should not be the same. In \cite{Setare:2015gss}, by explicit calculations it is shown that this is the case and the left and right central charges are inequal
\be\label{CC4} c_{L}=\frac{3\ell}{2G}\left[\left(\sigma+\frac{\gamma}{2\mu^2 \ell^2}+\frac{\eta}{2m^2\ell^2}\right)-\frac{1}{\mu\ell}\right],\quad  c_{R}=\frac{3\ell}{2G}\left[\left(\sigma+\frac{\gamma}{2\mu^2 \ell^2}+\frac{\eta}{2m^2\ell^2}\right)+\frac{1}{\mu\ell}\right].\ee
In fact, the inequality of $c_L$ and $c_R$ is due to the presence of the parity violating CS term in the action of GMMG theory, which originates from the diffeomorphism anomaly \cite{Kraus:2005zm} proportional to the difference of the central charges, i.e.
\be \label{hga} c_{L}-c_{R}=-\frac{3}{\mu G}.\ee
\subsubsection{The WAdS$_3$ black hole}
Another solution for the GMMG theory is the WAdS$_3$ black hole described by (\ref{wads}). The quasi-local conserved charges of this solution in CS like formulation were found in \cite{Setare:2017nlu}. The mass and angular momentum as two conserved charges are
\bea \label{mjwmmg} M&\!\!\!=\!\!\!&\frac{(\nu^2+3)}{16 \nu G}\,K_1\left(\nu(r_{+}+ r_{-})-\sqrt{(\nu^2+3) r_{+} r_{-}}\right)\,,\nn\\
 J&\!\!\!=\!\!\!&\frac{(\nu^2+3) \ell}{64\nu G}\left[\nu^2 K_2 \,(r_{+}- r_{-})^2-K_1\left(\nu(r_{+}+ r_{-})-\sqrt{(\nu^2+3) r_{+} r_{-}}\right)^2\right],\eea
where the coefficients $K_1$ and $K_2$ are defined as
\be \label{pmmg} K_1\equiv\sigma+\frac{\alpha(H_1+\ell^2 H_2)}{\mu}+\frac{F_1+\ell^2 F_2}{m^2}-\frac{\nu}{\mu \ell},\quad  K_2\equiv\sigma+\frac{\alpha(H_1+\ell^2 H_2)}{\mu}+\frac{F_1+\ell^2 F_2}{m^2}-\frac{\nu^2-3}{\mu \ell \nu}.\ee
 Also $H_1, H_2, F_1$, and $F_2$ are some constant parameters that have been introduced in \cite{Setare:2017nlu} to redefine 1-forms $f$ and $h$ in terms of $e$ and $J$ in CS like 3D gravities \cite{delPino:2015mna}. The black hole entropies of the outer and inner horizons are
\be \label{entmmg }S^{\pm}=\frac{\pi \ell}{4G} \left[K_1 \,\left(2\nu r_{\pm}-\sqrt{(\nu^2+3) r_{+} r_{-}}\right)+\frac{(\nu^2+3) (r_{\pm}- r_{\mp})}{2\mu\ell}\right]\,.\ee

Substituting from the relations (\ref{mjwmmg}) in the product of these entropies we have
 \be \label{prod50} S^{+}S^{-}=\frac{16\pi^2\nu^2\ell^2}{(\nu^2+3)^2}\left(1-\frac{K_2}{K_1}\right) M^2-\frac{4\pi^2\ell \nu K_2}{(\nu^2+3)G}\,J,\ee
it is obvious that the product depends explicitly on the mass and thus, it is not universal. As mentioned in section 2, the asymptotic symmetry group of WAdS$_3$ is denoted by the semi direct product of a Virasoro algebra with a U(1) current
algebra which only includes one of the central charges of the dual CFT. But as noted in \cite{Setare:2017nlu} for GMMG, and similarly \cite{Yekta:2016fif} in the case of NMG, we can enhance the symmetry to two versions of Virasoro algebras by using the Sugawara construction \cite{Sugawara:1967rw} with central charges
\be \label{CC20} c_{L}=\frac{6\ell \nu K_1}{(\nu^2+3)G},\quad c_{R}=\frac{6\ell \nu K_2}{(\nu^2+3)G}.\ee
Clearly the values of the left- and right-handed central charges are not equal which indicates the universality is not established and their differences remind us the holographic gravitational anomaly presented in (\ref{hga}).
\subsection{The exotic Einstein gravity}
Very long ago in the context of CS formalism, Witten proposed in \cite{Witten:1988hc} a counterpart of 3D Einstein gravity with the Lagrangian
\be \label{EG}
L_{EG}=\frac{1}{8\pi G}\left[\frac{\ell}{2} \left(\omega\cdot d\omega+\frac13 \omega\cdot\omega\times \omega \right)+\frac{1}{2\ell}\, e\cdot T\right],
\ee
where $T$ is the 2-form locally Lorentz covariant torsion as defined in (\ref{dtr}). This theory is known as the exotic Einstein gravity (EEG) that the full field equations (and not just their linearizations) are equivalent to 3D Einstein gravity, as is most easily seen from the fact that the action for both can be expressed as a linear combination of two $SL(2,R)$ CS actions \cite{Witten:1988hc}. It has been shown \cite{Carlip:1991zk}-\cite{Banados:1998dc} that the BTZ black hole is also a solution to the field equations of EEG models but with the fact that the roles of the mass and angular momentum are reversed. In order to clarify this assertion we rewrite the BTZ metric (\ref{BTZ}) by the following ADM functions
\be \label{func3} N^2=\frac{(r^2-r_{+}^2)(r^2-r_{-}^2)}{r^2 \ell^2}=-8Gm+\frac{r^2}{\ell^2}+\frac{16G^2j^2}{r^2}\,,\qquad N^{\varphi}=\frac{r_{+} r_{-}}{r^2 \ell}=-\frac{4Gj}{r^2}\,,\ee
where the mass, angular momentum and entropies of the inner and outer horizons for this solution are
\be \label{mjs}
M=m=\frac{r_+^2+r_-^2}{8g\ell^2}\,, \quad  J=j=\frac{r_+r_-}{4G\ell}\,, \quad S^{\pm}=\frac{\pi r_{\pm}}{2G}\,.
\ee
In the case of EEG, BTZ black holes are solutions for the equations of motion such that the mass and angular momentum are
\be \label{mjex}
M=\frac{j}{\ell}\,,\qquad J=\ell m\,,
\ee
that are also called ``exotic'' BTZ black holes. It is shown in \cite{Townsend:2013ela} that the entropy of the exotic BTZ black hole is proportional to the length ($2\pi r_-$) of the {\it inner} horizon, instead of the outer horizon. More precisely, the entropies of the outer and inner horizons in this case are
\be\label{eEEG} S^{\pm}=\frac{\pi r_{\mp}}{2G}.\ee

 Even though in the case of Einstein 3D gravity, it is shown \cite{Chen:2012mh} that the entropy product of the BTZ black hole is universal, but due to the relations (\ref{mjs}) and (\ref{mjex}) for the exotic black hole one finds
\be
S^+S^-=\frac{\pi^2\ell^2}{G}\,M\,,
\ee
which explicitly depends on the mass. In other words, the entropy product in this case is not universal and one expects \cite{Chen:2012mh, Chen:2013rb} that the left and right central charges of the dual CFT should not be the same. In fact it is discussed in \cite{Townsend:2013ela} that, due to the existence of parity odd CS term in the lagrangian (\ref{EG}) of this theory, the left and right central charges are $c_L=-c_R=\frac{3\ell}{2G}$, that confirms the statement of \cite{Chen:2012mh, Chen:2013rb}.
\subsection{The exotic general massive gravity}
 Just like to the previous sections we also consider an extension of NMG in this subsection which is known as general massive gravity (GMG) \cite{Bergshoeff:2009aq} but in the version of exotic, i.e. EGMG, that recently proposed in \cite{Ozkan:2018cxj}.
 The 3-form lagrangian of the EGMG theory is in the form
\bea
L_{\rm EGMG} &\!\!\!=\!\!\!& -\frac{\ell}{m^2}  \bigg[ f \cdot R(\omega) + \frac{1}{6m^4} f \cdot f \times f - \frac{1}{2m^2} f \cdot D(\omega) f + \frac{\nu}{2} f \cdot e \times e \nonumber\\
&\!\!\!-\!\!\!& \ m^2 h \cdot T(\omega) +\frac{\nu - m^2}{2} \left(\omega\cdot d\omega+\frac13 \omega\cdot\omega\times \omega \right)  + \frac{ \nu m^4}{3 \mu} e \cdot e \times e \bigg]\,,
\eea
with $\nu=\frac{1}{\ell^2}-\frac{m^4}{\mu^2}$. Further details of the theory and conventions could be tracked in \cite{Ozkan:2018cxj}. The BTZ black hole is a solution for EGMG theory with the mass and angular momentum \cite{Bergshoeff:2019rdb}
\be
M=-\frac{\ell m^2}{\mu} \tilde m+\left(1+ \frac{m^2}{\mu^2} - \frac{1}{m^2\ell^2} \right)\frac{\tilde j}{\ell}\,, \quad J=-\frac{\ell m^2}{\mu} \tilde j+ \ell^2 \left(1+ \frac{m^2}{\mu^2} - \frac{1}{ m^2\ell^2} \right)\tilde m\,,
\ee
where $\tilde m=\frac{r_+r_-}{4G\ell^2}$ and $\tilde j=\frac{r_+^2+r_-^2}{8G\ell}$ are the mass and angular momentum of the exotic BTZ black hole, respectively. Also the entropies of the outer and inner horizons are given by
\be
S^{\pm}=-\frac{\ell m^2}{\mu} \frac{\pi r_{\pm}}{2G}+\left(1+ \frac{m^2}{\mu^2} - \frac{1}{m^2\ell^2} \right)\frac{\pi r_{\mp}}{2G}.
\ee
At this stage, one can check the the entropy product of the EGMG BTZ solution takes to the form
\be
S^+S^-=\frac{\pi^2\ell}{G}\left[\left(1+ \frac{m^2}{\mu^2} - \frac{1}{m^2\ell^2} \right)\!J-\frac{m^2\ell^2}{\mu}M\right],
\ee
which is obviously mass dependent and according to \cite{Chen:2012mh, Chen:2013rb}, the left and right central charges should be different. In fact the explicit calculation \cite{Bergshoeff:2019rdb} shows that $c_L$ and $c_R$ are not the same
\be \label{CC30}
c_{L}=\frac{3\ell}{2G}\left[-\frac{\ell m^2}{\mu} + \left(1+ \frac{m^2}{\mu^2} - \frac{1}{m^2\ell^2} \right) \right], \quad c_{R}=\frac{3\ell}{2G}\left[-\frac{\ell m^2}{\mu} - \left(1+ \frac{m^2}{\mu^2} - \frac{1}{m^2\ell^2} \right) \right].
\ee

We have seen that in theories including the gravitational CS term the central charges of the dual CFT are not the same. In fact, due to the effects of gravitational anomalies on the boundary stress tensor, there are shifts in the central charges \cite{Kraus:2005zm}. On the other hand, we have shown that in these theories the $S^{+}S^{-}$ is not universal. Since in thermodynamics method $c_L=c_R$ is equivalent to the universality of the entropy product, we conclude that the CS term leads to mass dependence of the $S^+S^-$ and one is not allowed to employ the formula (\ref{prod100}). A similar behavior has already been observed in \cite{Kraus:2005vz} that in any diffeomorphism invariant Lagrangian $\mathcal{L}(R_{\mu\nu},g_{\mu\nu})$ of gravity, the central charges are given by ``$c\!-\!extremization$'' procedure, i.e. $c_{L}\!=\!c_{R}\!=\!\frac{\ell}{2G} \,g_{\mu\nu}\, \frac{\partial \mathcal{L}}{\partial R_{\mu\nu}}$, and the higher derivative terms constructed covariantly from curvature tensors and matter fields do not change this conclusion, but the only exceptions are CS terms.

It is known that the universality is a characteristic for a ``solution'', not for a ``gravity theory''. For instance, among the solutions of Einstein gravity, the $S^{+}S^{-}$ is universal for the Kerr black hole and for the Myers-Perry black hole in $D=5$, while it is mass dependent for Myers-Perry solutions in $D>5$. Another example can be found in the solutions of Einstein-$\Lambda$ gravity where $S^+S^-$ is universal in the case of BTZ black hole, while it is mass dependent for the Kerr-AdS solutions in $D \ge 4$\cite{Chen:2012mh}. But in this paper we encounter with a different case: we observed that the BTZ and WAdS$_3$ black holes respect the universality for diffeomorphism invariant theories in Secs. 2 and 3, while the $S^{+}S^{-}$ for these solutions is not universal in the theories including CS term. So, we expect the central charges not to be equal, as confirmed from the asymptotic symmetry group calculations.
\section{Summary and Conclusions}
In this paper, we studied the relation between the entropy product of the outer and inner horizons of 3D black hole solutions, with the central charges of their dual CFTs. Specifically, we studied the BTZ and warped AdS$_3$ black hole solutions of the NMG, as well as the charged black hole solution in CS Maxwell-Born-Infeld gravity. We observed that the entropy product of the inner and outer horizons is universal (mass independent) for these solutions.
According to the thermodynamics method introduced in \cite{Chen:2012mh,Chen:2012yd,Chen:2013rb}, this means that the central charges of left- and right-moving sectors in the dual CFT should be the same and one can also read the central charges form the entropy product as $c=6\, \frac{\partial}{\partial N} \left(\frac{S^{+}S^{-}}{4\pi^2}\right)$, where $N$ is a quantized charge of the solution like the angular momentum. We checked that the resulted central charges are in complete agreement with those who obtained by the other method such as asymptotic symmetry group, which means that the thermodynamics method of black hole/CFT correspondence also works truly in 3D gravity. We found the energies of left- and right-moving sectors  which together the central charges give the microscopic entropy consistent with the macroscopic one.

In addition, we investigated the entropy product for the BTZ and WAdS$_3$  black hole solutions in GMMG, EEG and EGMG theories. We found that the entropy product for these solutions are not universal. This means that the left and right central charges should not be the same according to the thermodynamics method. The inequality of $c_L$ and $ c_R$ for these solutions is confirmed when one calculates the central charges explicitly. It can also be interpreted due to the presence of CS term in the Lagrangians that often referred to as holographic gravitational anomaly.

\section*{Acknowledgment}
The authors would like to thank D. Grumiller for reading the manuscript and providing valuable comments.

   \end{document}